\documentclass[prl,aps,showpacs,superscriptaddress,onecolumn]{revtex4}

\usepackage{graphicx}
\usepackage{amssymb}
\usepackage{color}
\newcommand{\be}{\begin{equation}}
\newcommand{\ee}{\end{equation}}
\newcommand{\bea}{\begin{eqnarray}}
\newcommand{\eea}{\end{eqnarray}}
\newcommand{\ba}{\begin{array}}
\newcommand{\ea}{\end{array}}
\newcommand{\tr}{\mathrm{tr}}
\newcommand{\Tr}{\mathrm{Tr}}

\newcommand{\Rv}{R}
\newcommand{\bgamma}{\bar{\gamma}}
\newcommand{\Z}{\mathbb{Z}}
\newcommand{\change}[1]{\textcolor{black}{#1}}

\begin{document}
\title{Beta-functions of non-linear $\sigma$-models for disordered and interacting electron systems}
\author{Luca Dell'Anna}
\address
{Dipartimento di Fisica e Astronomia "Galileo Galilei" and CNISM, Universit\`a di Padova, Italy}
\begin{abstract}
We provide and study complete 
sets of one-loop renormalization group equations 
of several Finkel'stein non-linear $\sigma$ models, the effective 
field theories describing the diffusive quantum fluctuations in correlated disordered systems. 
We consider different cases according to the presence of certain symmetries induced by the original random 
Hamiltonians, and we show that, for interacting systems, the Cartan's classification of symmetry classes 
is not enough to uniquely determine their scaling behaviors.

\end{abstract}

\maketitle

\section{Introduction}

The characterization of disordered strongly interacting systems, in particular at low dimensions, is a formidable task in the whole history of modern condensed matter physics. A well-known effect of uncorrelated disorder is the Anderson localization, namely the exponential localization of waves in disordered media, predicted more than fifty years ago in the context of transport of non-interacting electrons in crystals. On the other hand, the interplay between disorder and interactions, instead, is a challenging problem still not fully understood. 
It is well known, in fact, that in the presence of impurities, in solid state systems, when the mean free path becomes of the same order or smaller than the de Broglie length of the particles, a ballistic description is not suitable for describing transport properties and one has to resort to quantum mechanics, treating the particles as waves. In such a situation, if time reversal symmetry is preserved, self-crossing quantum paths are crucial in generating quantum interference corrections which affect the probability for the traveling particles to reach the final target. This interference effect traduces to a reduction of conductivity in the system. If disorder becomes sufficiently strong one enter the so-called Anderson insulating regime where the wave packet of the traveling particles acquires an exponentially localized profile in real space \cite{anderson}. Moreover, in low dimensional systems, where quantum interferences are more effective, a small amount of disorder is enough to get Anderson localization \cite{mott}.
While these results are valid in non-interacting systems, a full understanding of the interplay of disorder and interactions, which is the origin of still unclear phenomena in condensed-matter physics, is still lacking, and represents an almost unexplored field. A theory for metal-insulator transition in strongly correlated and disordered systems for different universality classes is, so far, not completed. 
For instance, in two dimensions a small amount of disorder is enough to localize all the states, producing an Anderson insulator. By adding Coulomb interaction, instead, a metallic behavior may occur \cite{finkelstein}. However, if randomness does not break an additional symmetry, called chiral symmetry on a bipartite lattice, and in the absence of interaction, at the center of the band a delocalized state appears, producing a finite conductivity \cite{gade}. The situation is reversed by the presence of large momentum transfer interaction. In the latter case, besides spin density wave, another disorder-driven instability may arise, leading the system towards an insulating Anderson-Mott phase, where the system is both in a strong disorder and in a strong interaction regime \cite{npb, foster}. In two dimensions a metallic phase is, then, supposed to be unstable towards Anderson-Mott phase, while is stable in higher dimensions.
The combination of disorder and interactions can produce counter-intuitive effects as, for instance, the enhancement of the superconducting critical temperature $T_c$. As shown recently \cite{feigelman}, in the presence of a Cooper channel interaction, the critical $T_c$ is strongly enhanced, close to the Anderson transition. This phenomenon is induced by strong fluctuations and multifractal nature of the wavefunctions. A strong enhancement of $T_c$ by increasing disorder has been predicted also in two dimensions by means of the non-linear sigma model approach \cite{burmistrov} for the standard Anderson model (class AI), with short range interaction. 
\change{Enhancement of $T_c$ in a weak antilocalization regime has also been obtained \cite{burmistrov} for class AII with short-ranged interaction. 
Recently, it was showed \cite{prb} that the increase of $T_c$ driven by randomness can occur 
also for a chiral symmetry class \cite{zirnbauer,altland} (class BDI, for a random hopping model) where Anderson insulting phase is absent}. In the latter case other instabilities (charge density wave and antiferromagnetism) can occur, at temperatures much higher that those of clean systems. In other words, unexpectedly, disorder can generate a magnetic order \cite{prb}.

For that reason, it is important to carry on a complete analysis of the role of disorder combined with interactions, analyzing in great detail the effects of different types of random Hamiltonians which induce different symmetry properties in the system and different scaling behaviors. 
In order to study the role of randomness in electron systems, one can resort to a quantum field theory approach for disordered systems \cite{wegner,efetov}, further improved to deal with combined effects of interactions and disorder in terms of a renormalized Fermi liquid \cite{finkelstein,castellani,belitz}. The related Landau parameters acquires a scale dependence and, together with the conductance, form a full set of couplings of the so-called Finkel'stein non-linear $\sigma$-model, which flow under the action of the renormalization group (RG).
In this paper we will, therefore, present several sets of one-loop $\beta$-functions for all the couplings appearing in interacting non-linear $\sigma$-models corresponding to different correlated random Hamiltonians.  

\section{Model}
We will consider the following Hamiltonian on a two-dimensional 
square lattice made by free and interacting terms 
\be
{\cal H}={\cal H}_0+{\cal H}_I\,.
\ee
The first term is the single particle Hamiltonian, defined by 
\begin{eqnarray}
{\cal H}_0 &=& -\sum_{\langle ij\rangle} 
\sum_\sigma \left( t_{ij}{\rm e}^{i\phi_{ij}}
c^\dagger_{i\sigma}c^{\phantom{\dagger}}_{j\sigma} + h.c.\right)
-\sum_{i\sigma}\mu\,  c^\dagger_{i\sigma}c^{\phantom{\dagger}}_{i\sigma}
-\sum_{i}\vec{b}_i\cdot
c^\dagger_{i}\,\vec{\sigma}\,c^{\phantom{\dagger}}_{i}
\label{Hamiltonian}
\end{eqnarray}
where $\langle ij \rangle$ means that the sum is restricted to 
nearest neighbor sites, $c^\dagger_{i\sigma}$ creates an electron with 
spin $\sigma=\uparrow,\downarrow$ at site $i$, and 
$c^{\phantom{\dagger}}_{i\sigma}$ annihilates it, while, in the last term, 
$c^\dagger_i=\left(c^\dagger_{i\uparrow},\, c^\dagger_{i \downarrow}\right)$, and $c_i$ its conjugate, 
$\vec\sigma=(\sigma_x,\sigma_y,\sigma_z)$ 
Pauli matrices in the spin space, and $\vec{b}_i=(b_{x i},b_{y i},b_{z i})$ some local magnetic field. 
The hopping matrix elements $t_{ij}=t_{ji}$ are real, independent random gaussian 
variables, 
while $\phi_{ij}=-\phi_{ji}$ can be zero or not depending 
whether time reversal invariance holds or is broken. 
The spectrum of the Hamiltonian (\ref{Hamiltonian}), for $\mu=0$,  
possesses a nesting property. This implies an additional symmetry 
(called chiral or sublattice symmetry) at half filling \cite{gade,fabrizio}. 
The localization properties are quite different 
whether chiral symmetry holds, which corresponds to 
the Fermi energy $\mu=0$ (at half-filling and in the absence of 
on-site disorder), 
or broken, $\mu\neq 0$.  
In the latter case, the localization properties of   
(\ref{Hamiltonian}) are analogous to models in which   
on-site disorder is present 
($\sum_i\mu_i c^\dagger_{i\sigma}c^{\phantom{\dagger}}_{i\sigma}$) 
or next-nearest neighbor hopping is 
included, which break chiral symmetry everywhere in the spectrum. 
For this reason while dealing with a bipartite lattice that induces 
an higher degree of symmetry we can reduce the problem to the 
standard case only by introducing an on-site term in the 
Hamiltonian, thus spoiling chiral symmetry.
The last term in Eq.~(\ref{Hamiltonian}), instead, breaks the spin rotational symmetry, 
by a Zeeman term (with a uniform magnetic field $\vec{b}_i=\vec{b}\neq \vec{0}$) 
or in the presence of nonuniform field, as in the case of magnetic impurities. 

We now analyze the disordered Hamiltonian (\ref{Hamiltonian}) by using the 
replica trick method within the path integral formalism \cite{efetov}. 
We introduce the vector Grassmann variables  
$c_i$ and $\bar{c}_i$ with components 
$c_{i,\sigma,n,a}$ and 
$\bar{c}_{i,\sigma,n,a}$, where $i$ refers to a lattice site, $\sigma$ to the 
spin, \change{$n$ is the frequency index} 
and $a=1,\dots,n$ is the replica index, as well as the 
Nambu spinors 
\be
\label{nambu}
\Psi_i = \frac{1}{\sqrt{2}}
\left(
\begin{array}{c}
\bar{c}_i \\
i\sigma_y c_i\\
\end{array}
\right) \,\,\,\textrm{and }\,\,\,\bar{\Psi}_i=\left[C\Psi_i\right]^t,
\ee
with the charge conjugation matrix 
$C=i\sigma_y \tau_1$. Here and in the following, 
the Pauli 
matrices $\sigma_b$ ($b=x,y,z$) act on the spin components, 
and $\tau_b$ ($b=1,2,3$) on the Nambu components $\bar{c}$ and 
$c$. The action corresponding 
to (\ref{Hamiltonian}) is 
\begin{eqnarray}
S_0 &=& -\sum_{ij} \bar{\Psi}_i
\left( t_{ij}{\rm e}^{-i\phi_{ij}\tau_3} 
+\delta_{ij}\tau_3 \,\vec{b}_i\cdot \vec{\sigma} 
+\delta_{ij}\mu 
+i\delta_{ij} \hat\omega 
\right)
\Psi_j 
\label{S}
\end{eqnarray}
where the source term $\bar{\Psi}_i \hat\omega\Psi_i$ is introduced in order 
to reproduce positive and negative frequency propagators \cite{efetov}, 
\change{where $\hat\omega$ is a digonal matrix in frequency space}. 
If magnetic impurities are present, 
one has to consider the additional spin-flip scattering term $\sum_i \bar{\Psi}_i \tau_3 \,\vec{b}_i\cdot \vec{\sigma}\Psi_i$. 
The same term gives the Zeeman splitting in the 
presence of a uniform magnetic field $\vec{b}_i =\vec{b}$.
If an on-site term is present or if we are far from half filling, we 
have 
$\sum_i \mu_i\bar{\Psi}_i \Psi_i$, 
which spoils chiral symmetry, both for $\mu_i$ a random variable or $\mu_i=\mu\neq 0$ constant,  
respectively. 
Following the standard procedure, one integrate over the disorder obtaining an action which is quartic in the fermonic fields. By Hubbard-Stratonovich transformation one can decouple this term, introducing an auxiliary field $Q$. In this way one can integrate over the fermions getting an action which depends only on $Q$. The main contribution to the corresponding partition function is the solution of the saddle-point equation. Expanding around that vacuum state we get an effective theory describing the transverse charge fluctuations which has the form of a non-linear $\sigma$-model. 

On top of what briefly described so far we can consider also the presence of two-particle interactions, as originally done by Finkel'stein. We therefore include an interaction term to the action which, generally, can be written as follows 
\be
S_I=-
\sum_{ijlm}\sum_{\sigma\sigma'}U_{ijlm}\bar{c}_{i\sigma}\bar{c}_{j\sigma'}c_{l\sigma'}c_{m\sigma}
\ee
However it is more advantageous and transparent rewriting this term in Matsubara frequency and momentum spaces, splitting it into contributions from different phase-space regions,
\bea
S_I\simeq 
\frac{T}{2}
\sum_{|k|\ll k_F}\sum_{p_1 p_2 l n m} 
&&\Big\{
{\Gamma^0_s} \,\bar{c}_{n}(p_1)\,\sigma_0\,c_{n+l}(p_1+k)\,\bar{c}_{m}(p_2)\,\sigma_0\,c_{m-l}(p_2-k)
\\
\nonumber
&&
-{\Gamma^0_t} \,\bar{c}_{n}(p_1)\,\vec{\sigma}\,c_{n+l}(p_1+k)\,\bar{c}_{m}(p_2)\,\vec{\sigma}\,c_{m-l}(p_2-k)\\
\nonumber
&&+
{\Gamma^0_c} \sum_{\sigma,\sigma'=\uparrow,\downarrow}\,\bar{c}_{n\sigma}
(p_1)\,\bar{c}_{l-n,\sigma'}(k-p_1)\,c_{m\sigma'}(p_2)
\,c_{l -m,\sigma}(k-p_2)
\\
\nonumber
&&
+{\Gamma^3_s}\,\bar{c}_{n}(p_1)\,\sigma_0\,c_{n+l}(p_1+k+q_\pi)\,\bar{c}_{m}(p_2)\,\sigma_0\,c_{m-l}(p_2-k-q_\pi)
\\
\nonumber
&&
-{\Gamma^3_t} \,\bar{c}_{n}(p_1)\,\vec{\sigma}\,c_{n+l}(p_1+k+q_\pi)\,\bar{c}_{m}(p_2)\,\vec{\sigma}\,c_{m-l}(p_2-k-q_\pi)
\\
\nonumber
&&
+{\Gamma^3_c}\sum_{\sigma,\sigma'=\uparrow,\downarrow}\bar{c}_{n\sigma}
(p_1)\,\bar{c}_{l-n,\sigma'}(k-p_1+q_\pi)\,c_{m\sigma'}(p_2)
\,c_{l -m,\sigma}(k-p_2+q_\pi)\Big\}
\eea
where the term proportional to $\Gamma^0_s$ and $\Gamma^0_t$ are the singlet and triplet 
particle-hole interaction channels, describing small frequency-momentum transfer between 
a particle and a hole. 
The term proportional to $\Gamma^0_c$  is the 
particle-particle interaction channel or Cooper channel, and  
describes small frequency-momentum transfer between two particles or two holes.
The interaction amplitudes 
$\Gamma^0_{s,t,c}$ couple to the smooth local charge fluctuations while 
 $\Gamma^3_{s,t,c}$ to the staggered ones. In the presence of nesting property induced by 
the chiral symmetry, the staggered fluctuations become diffusive, therefore particle 
interactions with $q_{\pi}=(\pm \pi, \pm \pi)$ momentum transfer must be included in 
the effective action \cite{npb}.

\section{Finkel'stein non-linear $\sigma$-model}
In this section we will present directly the final effective action provided by the quantum field theory approach to the problem of impurities and interactions, referring to Refs.~\cite{finkelstein,npb,efetov,belitz,fabrizio} for more datails.  
The effective low-energy model which describe transverse charge fluctuations, 
derived in analogy with the original Finkel'stein model \cite{finkelstein}, is, then,  
the following \cite{npb}
\be
{S}[Q]={S}_{0}[Q]+{S}_{I}[Q]\,,
\label{act}
\ee
where ${S}_0[Q]$ is the non-interacting part
\be
{S}_{0}[Q]=\frac{\pi}{32}
\int d\Rv\, \left\{\sigma\, \Tr\left(\vec{\nabla}Q \cdot\vec{\nabla} 
Q^\dagger\right)
-{8\nu\, z}\,
\Tr\left(\hat\omega Q\right)
-\frac{\Pi}{8}\,
\Tr\left(Q^\dagger\vec{\nabla}Q\rho_3\right)
\cdot \Tr\left(Q^\dagger\vec{\nabla}Q\rho_3\right)
\right\},
\label{NLsM}
\ee
and ${S}_I[Q]$ the contribution from $e$-$e$ interactions
\bea
\nonumber {S}_{I}[Q]=\frac{\pi^2\nu^2}{32}\hspace{-0.1cm}
\int^{\prime}
\hspace{-0.1cm}
\Big\{
\Gamma^{\alpha}_t
\sum_{\beta=0,3}\tr(Q^{rr}_{n,n+p} 
\rho_\alpha\,\tau_\beta\,\vec{\sigma})
\cdot\tr(Q^{rr}_{\ell+p,\ell}\rho_\alpha\, \tau_\beta\, \vec{\sigma})
-\Gamma^{\alpha}_s
\sum_{\beta=0,3}\tr(Q^{rr}_{n,n+p} \rho_\alpha\, \tau_\beta\, \sigma_0)
\,\tr(Q^{rr}_{\ell+p,\ell}\rho_\alpha\, \tau_\beta\, \sigma_0)
\\+\Gamma^{\alpha}_c
\sum_{\beta=1,2}\tr(Q^{rr}_{n+p,-n} \rho_\alpha\, \tau_\beta\, \sigma_0)
\,\tr(Q^{rr}_{\ell+p,-\ell} \rho_\alpha\, \tau_\beta\, \sigma_0)
\Big\}
\label{inter}
\eea
The symbol $\int^\prime$ in Eq.~(\ref{inter}) means an integral over real
space,  $\int d\Rv$, and a sum over smooth ($\alpha=0$) and staggered 
($\alpha=3$) modes, $n_r$ replicas and
Matsubara frequencies, i.e.
$\int^\prime\equiv \int d\Rv\sum_{\alpha=0,3}\sum_r                        
\sum_{\ell,n,p}$, where $r$ is the replica index and $\ell, n, p$ are
Matsubara indices.
The matrix field $Q$ is constrained by the condition $QQ^\dagger=\mathbb{I}$.
The coupling $\sigma$
corresponds to the Kubo formula for the charge conductivity at the Born level;
$\nu$ is the density of states at the Fermi energy at the Born approximation;
$\Gamma^0_s, \Gamma^0_t, \Gamma^0_c$ and
$\Gamma^3_s, \Gamma^3_t, \Gamma^3_c$ are related to the Landau scattering 
amplitudes \cite{castellani,belitz} or to the interaction parameters of a bipartite 
Hubbard-like model \cite{foster},  
for smooth ($\Gamma^0$), and 
staggered sublattice ($\Gamma^3$) 
components, in the particle-hole singlet, particle-hole triplet 
and particle-particle Cooper channels, respectively; $z$ is a 
renormalization constant; $\hat\omega$ is a diagonal matrix made of Matsubara
frequencies. The last term in Eq.~(\ref{NLsM}) is the anomalous additional
term which is present only if the sublattice symmetry is preserved
\cite{gade} and the coupling $\Pi$ is
related to the staggered density of states fluctuations \cite{fabrizio}, 
associated with quenched orientational fluctuations of bond strength dimerization \cite{foster}.
$\tau_1, \tau_2, \tau_3$ are Pauli matrices in particle-hole space;
$\sigma_x, \sigma_y, \sigma_z$ are Pauli matrices in spin space; $\rho_3$ is 
the third Pauli matrix in the sublattice space;
$\tau_0, \sigma_0, \rho_0$ are identity matrices in the corresponding spaces.
The trace ``$\Tr$'' is made over all spaces (particle-hole, spin, sublattice,
replica and Matsubara spaces), while the trace ``$\tr$''
is over particle-hole, spin and sublattice spaces.
The action in Eqs.~(\ref{act})-(\ref{inter}) is tailored to describe
two-sublattice models, namely when sublattice symmetry is preserved and
staggered modes are massless.
When the sublattice symmetry is broken, staggered modes become massive, then
the last term in
Eq.~(\ref{NLsM}) 
and the terms in Eq.~(\ref{inter}) with
$\alpha=3$ should be put to zero ($\Pi=\Gamma_s^3=\Gamma_t^3=\Gamma_c^3=0$).
In this way one recovers the standard
Finkel'stein non-linear $\sigma$-model \cite{finkelstein}, provided that
$Q$ takes values in the proper coset space (see Table \ref{Table}). 
If, instead, staggered modes are massless, those interacting terms are
naturally generated by the RG flow.\\
In order to get rid of the \change{parameter $z$}, it is usually
convenient to define, together with the disorder parameters 
\be
g= 1/(2\pi^2 \sigma), \;\;\;
\Gamma= \Pi/(\sigma+n_r\Pi),
\ee
the following interacting parameters, in the smooth interacting channels
\be
\label{gamma0}
\gamma_s= 2\nu\Gamma^0_s/z,\;\;\;
\gamma_t= 2\nu\Gamma^0_t/z,\;\;\;
\gamma_c= 2\nu\Gamma^0_c/z,
\ee
and in the staggered ones
\be
\label{gamma3}
\bgamma_s= 2\nu\Gamma^3_s/z,\;\;\;
\bgamma_t= 2\nu\Gamma^3_t/z,\;\;\;
\bgamma_c= 2\nu\Gamma^3_c/z.
\ee

\begin{table}[!h]

\begin{tabular}{|c||c|c|c||c||c|}
 \hline

\#&${\cal S}$ &$\cal{T}$ &SU$(2)$& Coset space &  Class \\ 
 \hline\hline

$1)$&$\mu=0$ & $\phi=0$ & $\vec{b}_i=0$ & U$(8n)$/Sp$(4n)$
& BDI \\  \hline

$2)$&$\mu=0$&$\phi\neq 0$ & $\vec{b}_i=0$ & U$(4n)$$\times$U$(4n)$/U$(4n)$
& AIII \\  \hline

$3)$&$\mu=0$&$\phi\neq 0$ &$\vec{b}_i=\vec{b}$ & U$(4n)$/U$(2n)$$\times$U$(2n)$ & A \\  \hline

$4)$&$\mu=0$&$\phi=0$ &$\vec{b}_i\neq \vec{b}$ & Sp$(2n)$/U$(2n)$
& C \\  \hline

$5)$&$\mu= 0$&$\phi\neq0$ &$\vec{b}_i\neq \vec{b}$ & U$(2n)$/U$(n)$$\times$U$(n)$&A\\ \hline

$6)$&$\mu\neq 0$& $\phi=0$& $\vec{b}_i=0$ & Sp$(4n)$/Sp$(2n)$$\times$Sp$(2n)$
& AI \\  \hline

$7)$&$\mu\neq 0$&$\phi\neq 0$ & $\vec{b}_i=0$ & U$(4n)$/U$(2n)$$\times$U$(2n)$ 
& A \\  \hline

$8)$&$\mu\neq 0$&$\phi\neq 0$ & $\vec{b}_i=\vec{b}$ & U$(2n)$$\times$U$(2n)$/U$(2n)$
& AIII \\  \hline

$9)$&$\mu\neq 0$&$\phi=0$ &$\vec{b}_i\neq \vec{b}$ & U$(2n)$/U$(n)$$\times$U$(n)$&A \\ \hline

$10)$&$\mu\neq 0$&$\phi\neq0$ &$\vec{b}_i\neq \vec{b}$ & U$(2n)$/U$(n)$$\times$U$(n)$&A
\\  \hline
\end{tabular}
\caption{The $\sigma$-model target manifolds \change{(coset spaces)} corresponding to different realizations of ${\cal H}_0$, Eq.~(\ref{Hamiltonian}), and their corresponding symmetry classes \change{of these non-interacting random Hamiltonians} according to the Cartan's classification, 
and where chiral sublattice symmetry, ${\cal S}$, is preserved 
($\mu=0$) or broken ($\mu\neq 0$), or time reversal symmetry, ${\cal T}$, holds ($\phi=0$) or not ($\phi\neq 0$), and in the presence of spin rotational symmetry, SU$(2)$, ($\vec{b}_i=0$, $\forall i$), which can be lowered by a Zeeman field ($\vec{b}_i=\vec{b}$, $\forall i$) or completely spoiled by a nonuniform magnetic field, as in the presence of magnetic impurities, ($\vec{b}_i\neq \vec{b}$). The number  $n$ for the coset spaces, means $n=$ \# of replicas $\times$ $\#$ of positive Matsubara frequencies.}
\label{Table}
\end{table}

\section{Perturbative $\beta$-functions for different symmetries}

We now consider the scaling behavior of our action, for all the cases listed 
in Table~\ref{Table}, by means of the 
Wilson-Polyakov renormalization group approach \cite{wilson,polyakov}, 
introducing slow and fast modes 
\be
\label{UQU}
Q=\tilde U_s^\dagger Q_f U_s=\tilde U_s^\dagger\tilde U_f^\dagger Q_{sp} U_fU_s
\ee
where $U_{s,f}$ are unitary transformations involving transverse massless 
fluctuations in the slow and fast energy sectors, 
and $\tilde U^\dagger=CU^t C^t=\rho_1U^\dagger\rho_1=\rho_2 U^\dagger\rho_2$. 
If sublattice symmetry is broken  $\tilde U=U$. 
$Q_{sp}$ is the saddle point solution. Introducing an ultraviolet energy 
cutoff $\Lambda$, we can integrate over the fast modes which are defined for 
energies $<\Lambda/s$, with $s\ge 1$ the scaling factor (at finite temperature 
this scaling factor can be $s=\Lambda/T$). The details of the calculation can be found in \cite{npb} \change{and in the Appendix}. The scaling behavior of 
the couplings of the interacting $\sigma$-models are described by the so-called $\beta$-functions, $\beta_{g}=dg/d\ell$, $\beta_\Gamma=d\Gamma/d\ell$, $\beta_\gamma=d\gamma/d\ell$, 
in the zero replica limit $n_r\rightarrow 0$, where $\ell=\ln(s)$ is the scaling parameter. 
Perturbative $\beta$ functions for (non-interacting) non-liner $\sigma$-models on all types of symmetric 
spaces were calculated long ago up to four-loop order \cite{hikami,wegner89}. Here we present one-loop $\beta$ functions for (interacting) Finkel'stein non-linear $\sigma$-models at all orders in the interacting parameters, \change{except those which can lead to instabilities}. 

\subsection{Case 1) class BDI, with ${\cal S}$, ${\cal T}$, SU$(2)$} 
Let us first consider the case in which sublattice symmetry, time reversal symmetry and spin rotational invariance are preserved. In this case, which 
posseses the largest symmetry among the cases we shall be considering, 
the one-loop RG equations for the full set of coupligs 
\change{(at all orders in the 
interaction parameters $\gamma_s$, $\gamma_t$, $\bar\gamma_c$ and to leading orders in $\bar\gamma_s$, $\bar\gamma_t$, $\gamma_c$)}, 
at $d=2+\epsilon$ dimensions, are the following
\begin{eqnarray}
\hspace{-0.cm}\frac{d g}{d\ell} &=& -\epsilon g+
{g^2}\left\{6+\frac{1-\gamma_s}{\gamma_s}\,\ln(1-\gamma_s)-\frac{1}{2}\bgamma_s-3\,\frac{1+\gamma_t}{\gamma_t}\,\ln(1+\gamma_t)+\frac{3}{2}\bgamma_t-\gamma_c 
+ 2\,\frac{1-\bgamma_c}{\bgamma_c}\,\ln(1-\bgamma_c)\right\}
\label{(g)}
\\
\frac{d\Gamma}{d\ell}&=&
\epsilon \Gamma +8g +\frac{\Gamma}{g}\frac{dg}{d\ell}\\
\frac{d\gamma_t}{d\ell}&=&
g\left\{\left(1+\gamma_t\right)
\left(-\frac{3}{2}-\frac{\Gamma}{8}+\frac{\gamma_s}{2}-\frac{\bgamma_s}{2}-\gamma_c+\bgamma_c+\frac{3}{2}\bgamma_t\right)
+\left(1+\gamma_t\right)^2\left(\frac{3}{2}+\frac{\Gamma}{8}+\bgamma_s-\bgamma_t\left(1-\frac{\Gamma}{4}\right)+2\gamma_c\right)
\right\}\\
\nonumber \frac{d\bgamma_t}{d\ell}&=&g\left\{
\left(\frac{1}{2}+\frac{\Gamma}{4}\right)\gamma_t+\left(\frac{3}{2}+\frac{\Gamma}{8}+\ln\left[\frac{(1-\bgamma_c)^2(1-\gamma_s)}{1+\gamma_t}\right]\right)\bgamma_t+
\frac{\bgamma_s}{2}+\frac{\gamma_s}{2}+\gamma_c+\bgamma_c\right.\\
\label{(bt)}
&&+\bgamma_t\left(\frac{1}{2}\gamma_s-\frac{3}{2}\gamma_t+\bgamma_c
\left.+\gamma_c-\frac{3}{2}\bgamma_t+\frac{1}{2}\bgamma_s\right)\right\}+(\bgamma_t)^2\\
\frac{d\gamma_s}{d\ell}&=&
g\left\{\left(1-\gamma_s\right)
\left(\frac{1}{2}+\frac{\Gamma}{8}+\frac{\bgamma_s}{2}+\gamma_c-\bgamma_c+\frac{3}{2}\gamma_t-\frac{3}{2}\bgamma_t\right)
-\left(1-\gamma_s\right)^2\left(\frac{1}{2}+\frac{\Gamma}{8}+\bgamma_s\left(1-\frac{\Gamma}{4}\right)-\frac{2\bgamma_c}{1-\bgamma_c}-3\bgamma_t\right)
\right\}\phantom{-.}\\
\nonumber \frac{d\bgamma_s}{d\ell}&=&g\left\{
\frac{3}{2}\gamma_t+\frac{3}{2}\bgamma_t+\left(\frac{1}{2}+\frac{\Gamma}{8}+\ln\left[(1+\gamma_t)^3(1-\gamma_s)(1-\bgamma_c)^2\right]\right)\bgamma_s-\left(\frac{1}{2}-\frac{\Gamma}{4}\right)\gamma_s+\bgamma_c\right.\\
\label{(bs)}
&&+\gamma_c\left(1-4\ln(1-\bgamma_c)\right)
+\bgamma_s\left(\frac{1}{2}\gamma_s-\frac{3}{2}\gamma_t+\bgamma_c
\left.-\frac{3}{2}\bgamma_t+\frac{1}{2}\bgamma_s+\gamma_c\right)\right\}-(\bgamma_s)^2 \phantom{--}\\
\nonumber \frac{d\gamma_c}{d\ell}&=&g\left\{
\frac{3}{2}\gamma_t+\frac{3}{2}\bgamma_t+\frac{\gamma_s}{2}+\frac{\bgamma_s}{2}+\left(1+\frac{\Gamma}{8}+\ln\left[\frac{(1+\gamma_t)^3(1-\bgamma_c)^2}{(1-\gamma_s)}\right]\right)\gamma_c+\frac{\Gamma}{4}\bgamma_c\right.\\
\label{(c)}
&&-2\bgamma_s\ln(1-\bgamma_c)
+\gamma_c\left(\frac{1}{2}\gamma_s-\frac{3}{2}\gamma_t+\bgamma_c
\left.-\frac{3}{2}\bgamma_t+\frac{1}{2}\bgamma_s+\gamma_c\right)\right\}-(\gamma_c)^2\\
\nonumber \frac{d\bgamma_c}{d\ell}&=&g\left\{-1-\gamma_c+\gamma_s+\bgamma_s
-2\bgamma_c\left(\ln(1-\bgamma_c)-\frac{\bgamma_c}{\gamma_s}\ln(1-\gamma_s)\right)
\right.\\
&&+\left(1-\bgamma_c\right)\left(3+\frac{\Gamma}{8}+3\gamma_c-\frac{\gamma_s}{2}-\frac{3}{2}\bgamma_s+\frac{3}{2}\left(\gamma_t-\bgamma_t\right)\right)
\left.\hspace{-0.08cm}+\left(1-\bgamma_c\right)^2
\hspace{-0.08cm}
\left(\bgamma_s+3\bgamma_t-2-\frac{\Gamma}{8}-2\gamma_c+\frac{\Gamma}{4}\gamma_c\right)\right\}
\end{eqnarray}
For the sake of completeness we report the scaling equation for the density of states (DOS),  
which is
\be
\frac{d\nu}{d\ell}=g\nu\left\{\frac{1}{2}\ln(1-\gamma_s)+\frac{3}{2}\ln(1+\gamma_t)+\ln(1-\bgamma_c)
-\frac{3}{2}\bgamma_t+\frac{\bgamma_s}{2}+\gamma_c
+1+\frac{\Gamma}{8}\right\}\,.
\label{nu}
\ee
The scaling behavior of $z$, appearing in the second term of Eq.~(\ref{NLsM}) can be obtained from 
Eq.~(\ref{nu}) keeping only the linear terms in the interaction parameters since 
in the calculation for the renormalization of $z$ it turns out that higher order terms 
cancel out exactly \cite{npb}. This observation for $z$ is valid also for the other cases which follow. 
In this case
\be
\frac{dz}{d\ell}=gz\left(\frac{3\gamma_t}{2}-\frac{\gamma_s}{2}-\bgamma_c-\frac{3\bgamma_t}{2}+\frac{\bgamma_s}{2}+\gamma_c+1+\frac{\Gamma}{8}\right).
\ee 
The equations above are symmetric under the transformation 
$\gamma_s=\bgamma_c\leftrightarrow -\gamma_t$ and  
$\bgamma_s=\gamma_c\leftrightarrow -\bgamma_t$. 
This symmetry property of the parameters can be obtained by 
particle-hole transformation of the original fermionic fields defined on the lattice, 
$c_{i\uparrow} \rightarrow c_{i\uparrow}$, $c_{i\downarrow} \rightarrow (-)^ic^\dagger_{i\downarrow}$, which maps charge to spin and vice versa. As a result, by imposing 
\bea
\label{gamma}
\gamma_s=\bgamma_c=-\gamma_t\\
\bgamma_s=\gamma_c=-\bgamma_t
\label{bgamma}
\eea
we can define a subspace of the full space of parameters, invariant under RG flow. 
For $\epsilon=0$ and in the limit of all $\gamma\rightarrow 0$, 
Eq.~(\ref{(g)}) becomes 
$dg/d\ell=0$, namely, the resistance $g$ is scale invariant. This non-interacting behavior
is what is called the Gade-Wegner criticality \cite{gade,evers}, since it is valid at all loop orders. 
The last terms in Eqs.~(\ref{(bt)}), (\ref{(bs)}), and (\ref{(c)}), those not
coupled to $g$, are actually the terms which can drive the
system to antiferromagnetic spin density wave ($\bgamma_t$), charge density 
wave ($\bgamma_s$), and $s$-wave superconductivity ($\gamma_c$) also in clean
systems, obtained by ladder resummations. 
In the presence of interactions, and far from instabilities, the combined effects of disorder and interactions can drive the system towards an insulating Anderson-Mott phase, where the system is both in a strong disorder (maily the parameter $\Gamma$ increases) and in a strong interacting regime \cite{npb}, as in the spinless case with broken time reversal symmetry \cite{foster}. In two dimensions a metallic phase 
is, then, supposed to be unstable towards Anderson-Mott phase, while should be stable in higher dimensions.  
These equations have been extensively studied \cite{prb} when 
we allow the system to undergo some instabilities 
(as for example, by using initial conditions $\bgamma_t>0$, $\bgamma_s<0$, or $\gamma_c<0$). 
In this cases, unexpectedly, disorder can even promote such instabilities 
driving the system towards antiferromagnet, charge density or superconducting 
phases \cite{prb}.

\subsection{Case 2) class AIII, with ${\cal S}$ and SU$(2)$, without ${\cal T}$} 
If we breaks time reversal symmetry the complete one-loop $\beta$-functions are the following
\begin{eqnarray}
\label{gAIII}
\frac{d g}{ d\ell} &=&-\epsilon g+g^2\left\{4-3\frac{1+\gamma_t}{\gamma_t}\ln(1+\gamma_t)+\frac{3}{2}\bgamma_t-\frac{\bgamma_s}{2}+\frac{1-\gamma_s}{\gamma_s}\ln(1-\gamma_s)\right\}\\
\label{GammaAIII}
\frac{d\Gamma}{d\ell}&=&
\epsilon \Gamma +4g +\frac{\Gamma}{g}\frac{dg}{d\ell}\\
\frac{d\gamma_t}{d\ell}&=&g\left\{
\left(1+\gamma_t\right)\left(-\frac{1}{2}-\frac{\Gamma}{8}+\frac{\gamma_s}{2}-\frac{\bgamma_s}{2}+\frac{3}{2}\bgamma_t\right)
+\left(1+\gamma_t\right)^2\left(\frac{1}{2}+\frac{\Gamma}{8}+\bgamma_s-\bgamma_t+\frac{\Gamma}{4}\bgamma_t\right)
\right\}\\
\nonumber \frac{d\bgamma_t}{d\ell}&=&g\left\{
\left(\frac{1}{2}+\frac{\Gamma}{4}\right)\gamma_t+\left(\frac{1}{2}+\frac{\Gamma}{8}-\ln(1+\gamma_t)+\ln(1-\gamma_s)\right)\bgamma_t+
\frac{\bgamma_s}{2}+\frac{\gamma_s}{2}\right.\\
&&\left.+\,\bgamma_t\left(\frac{1}{2}\gamma_s-\frac{3}{2}\gamma_t
 -\frac{3}{2}\bgamma_t+\frac{1}{2}\bgamma_s\right)\right\}
+(\bgamma_t)^2\\
\frac{d\gamma_s}{d\ell}&=&g\left\{
\left(1-\gamma_s\right)\left(-\frac{1}{2}+\frac{\Gamma}{8}+\frac{\bgamma_s}{2}+\frac{3}{2}\gamma_t-\frac{3}{2}\bgamma_t\right)
+\left(1-\gamma_s\right)^2\left(\frac{1}{2}-\frac{\Gamma}{8}-\bgamma_s+\frac{\Gamma}{4}\bgamma_s+3\bgamma_t\right)
\right\}\\
\nonumber \frac{d\bgamma_s}{d\ell}&=&g\left\{
\frac{3}{2}\left(\gamma_t+\bgamma_t\right)+\left(\frac{\Gamma}{8}-\frac{1}{2}+\ln(1-\gamma_s)+3\ln(1+\gamma_t)\right)\bgamma_s-\left(\frac{1}{2}-\frac{\Gamma}{4}\right)\gamma_s\right.\\
&&\left.+\,\bgamma_s\left(\frac{1}{2}\gamma_s-\frac{3}{2}\gamma_t
-\frac{3}{2}\bgamma_t+\frac{1}{2}\bgamma_s
\right)\right\}-(\bgamma_s)^2
\end{eqnarray}
Also in this case, for completeness, we report the $\beta$-functions for the DOS and $z$
\bea
\frac{d\nu}{d\ell}&=&g\nu\left\{\frac{1}{2}\ln(1-\gamma_s)+
\frac{3}{2}\ln(1+\gamma_t)-\frac{3}{2}\bgamma_t+\frac{1}{2}\bgamma_s+\frac{\Gamma}{8}\right\}\\
\frac{dz}{d\ell}&=&gz\left(\frac{3\gamma_t}{2}-\frac{\gamma_s}{2}-
\frac{3\bgamma_t}{2}+\frac{\bgamma_s}{2}+\gamma_c+\frac{\Gamma}{8}\right).
\eea
These equations can be mapped to those for disordered $d$-wave superconductors in the presence of both 
chiral symmetry and time reversal invariance \cite{npb}, by the simple substitutions  
\begin{eqnarray*}
\gamma_s\rightarrow \bgamma_c\\
\bgamma_s\rightarrow \gamma_c
\end{eqnarray*}
This accidental equivalence is 
consistent with the fact that both systems belong to the same symmetry class.
However, this is a coincidence since, generally, interacting systems which belong to the same symmetry classes can have different $\beta$-functions (see, for instance, cases 3) and 7) below).  
\begin{figure}[ht!]
\includegraphics[width=7cm]{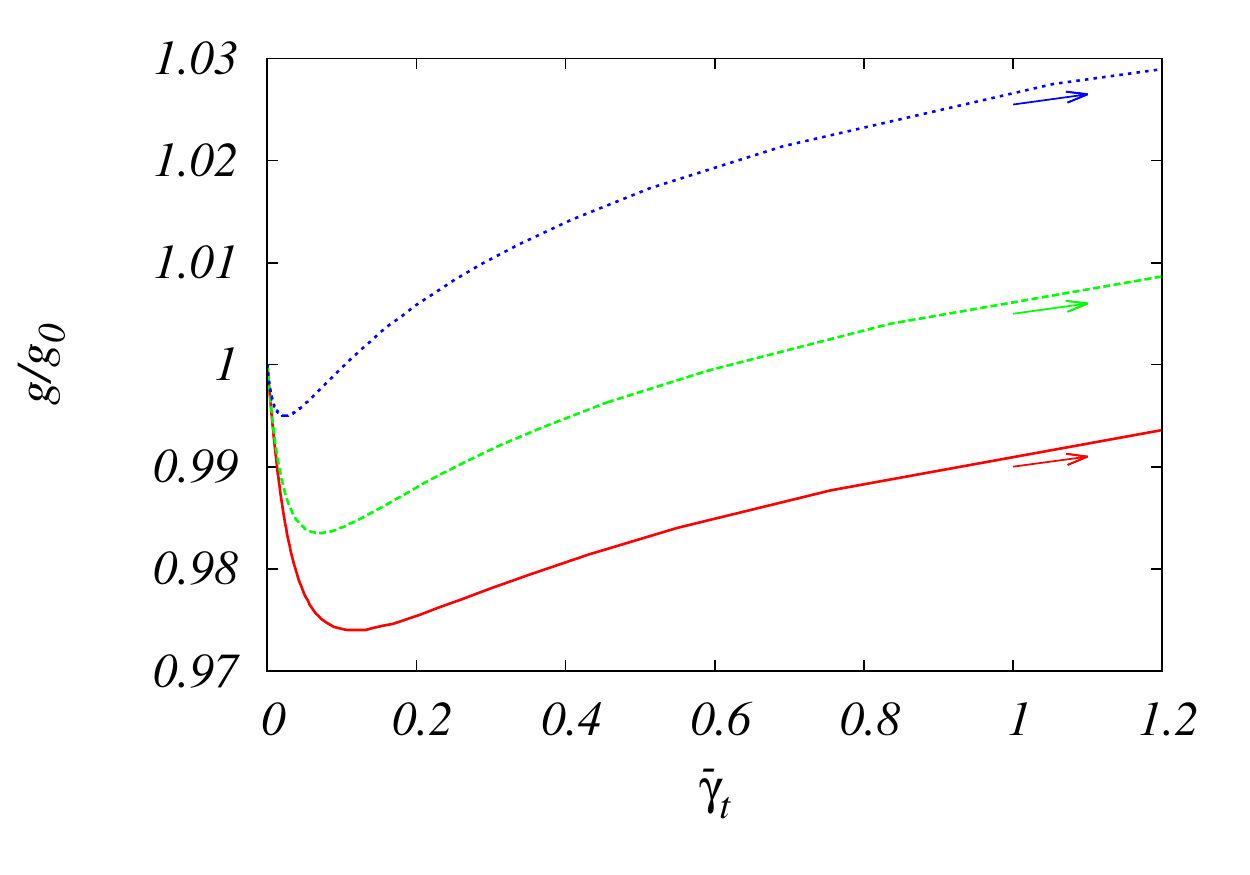}
\caption{RG flow lines for class AIII in case 2), on the $g$-$\bgamma_t$ plane, for $\epsilon=0$, 
with initial values $g_0=g(\ell=0)=0.01$, $\bgamma_{t}(0)
=\bgamma_{s}(0)=0$, $\Gamma(0)=0$, and $\gamma_{s}(0)=\gamma_{t}(0)=0.01$ (blue dotted line), $0.05$ 
(green dashed line), $0.1$ (red solid line).}
\label{case2.fig}
\end{figure}
By numerically solving the RG equations for $\epsilon=0$, one can find that, before reaching the strong coupling regime for the resistance $g$, the system enter the antiferromagnetic phase where $\bgamma_t$ diverges. In Fig.~\ref{case2.fig} we can see that $g$ first slightly decreases, mainly due to the effect of smooth spin fluctuations, but then increases again, anyway not fast enough for the staggered spin fluctuations which drive the system to a N\'eel state.

\subsection{Case 3) class A: with ${\cal S}$ and a uniform magnetic field (${\cal T}$ and SU$(2)$ broken)}

If we now break not only the time reversal symmetry but also the spin rotational invariance, by 
applying, for instance, a external uniform magnetic field which induces a Zeeman splitting, then the $\beta$-functions become 
\begin{eqnarray}
\frac{d g}{ d\ell}
&=&-\epsilon g+
g^2\left\{2+\frac{1-\gamma_s}{\gamma_s}\ln(1-\gamma_s)
-\frac{1+\gamma_t}{\gamma_t}\ln(1+\gamma_t)+\bgamma_t\right\}\\
\frac{d\gamma_t}{d\ell}&=&g\left\{\frac{1}{2}
\left(1+\gamma_t\right)\left(1+\gamma_s+2\bgamma_t\right)-\frac{1}{2}\left(1+\gamma_t\right)^2
\right\}\\
\frac{d\bgamma_t}{d\ell}&=&g\left\{
\frac{\gamma_t}{2}-\left(3\ln(1+\gamma_t)+\ln(1-\gamma_s)\right)\bgamma_t-\frac{1}{2}\gamma_t\bgamma_t+\frac{1}{2}\gamma_s\bgamma_t-\bgamma_t\bgamma_t\right\}+(\bgamma_t)^2  \\
\frac{d\gamma_s}{d\ell}&=&g\left\{\frac{1}{2}
\left(1-\gamma_s\right)\left(\gamma_t-2\bgamma_t-1\right)
+\frac{1}{2}\left(1-\gamma_s\right)^2\left(1+4\bgamma_t\right)\right\}
\end{eqnarray}
and the scaling behavior of the DOS and $z$ are ruled by the following 
equations  
\bea
\frac{d\nu}{d\ell}&=&g\nu\left\{\frac{1}{2}\ln(1-\gamma_s)+
\frac{1}{2}\ln(1+\gamma_t)-\bgamma_t\right\}\\
\frac{dz}{d\ell}&=&gz\left(\frac{\gamma_t}{2}-\frac{\gamma_s}{2}-
{\bgamma_t}\right).
\eea
Notice that this case belongs to the same symmetry class of the integer quantum Hall effect (IQHE). 
In the presence of a topological $\theta$-term in the action (see below), the system has a critical point 
in the strong coupling regime \cite{pruisken,khmelnitskii}. 
However, in the presence of interactions, under renormalization group procedure, starting from weak coupling, 
the system preferably flows towards an antiferromagnetic order before exploring the strong (in $g$) coupling 
regime. This behavior is analogous to case 2) except from the fact that smooth spin fluctations here is much 
less effective due to the lowering of the SU$(2)$ symmetry.

\subsection{Case 4) class C, with ${\cal S}$ and ${\cal T}$, magnetic impurities (SU$(2)$ broken)}
\begin{figure}[ht!]
\includegraphics[width=7cm]{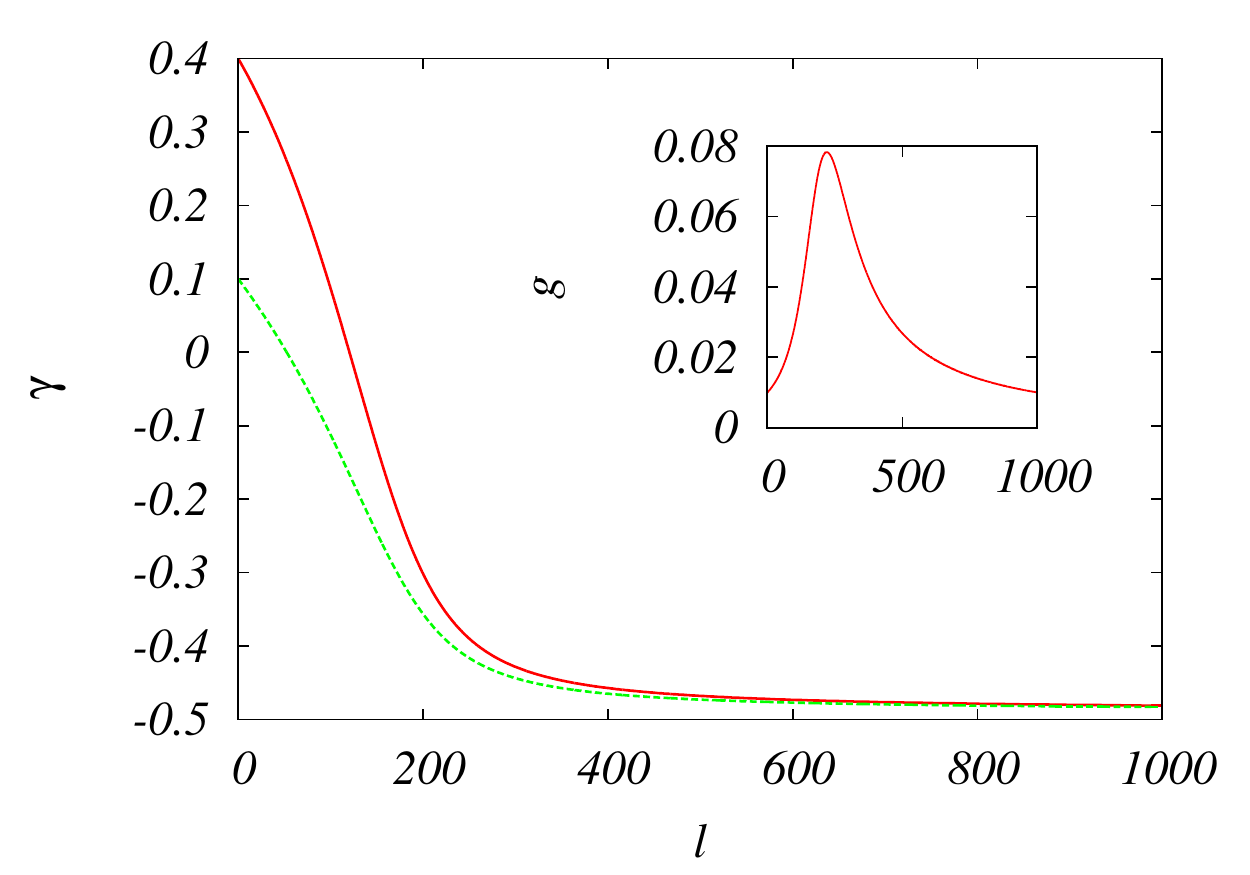}
\caption{Main plot: $\gamma_s(\ell)$ (solid red line) and $\bgamma_c(\ell)$ (dashed green line) 
for class C, case 4), as functions of $\ell$, the scaling parameter, 
with initial values $g(0)=0.01$, $\gamma_{s}(0)=0.4$, $\bgamma_{c}(0)=0.1$. 
In the inset: $g$ as a function of $\ell$, for the same initial values.}
\label{case4.fig}
\end{figure}

In the presence of magnetic impurities the spin-triplet modes become massive while the particle-particle 
Cooper interaction survives only in the staggered channel, so that we get  

\begin{eqnarray}
\frac{dg}{d\ell} 
&=&-\epsilon g+g^2\left\{\frac{7}{2}+\frac{1-\gamma_s}{\gamma_s}\ln(1-\gamma_s)+2\frac{1-\bgamma_c}{\bgamma_c}\ln(1-\bgamma_c)\right\}\\
\frac{d\gamma_s}{d\ell} 
&=&g\left\{-\left(1-\gamma_s\right)\left(1+\bgamma_c\right)+\left(1-\gamma_s\right)^2\left(\frac{1+\bgamma_c}{1-\bgamma_c}\right)
\right\}\\
\frac{d\bgamma_c }{d\ell} 
&=&g\left\{\frac{1}{2}\left(\bgamma_c-\gamma_s\right)\left(1+\bgamma_c\right)-2\bgamma_c\ln(1-\bgamma_c)+2\frac{\bgamma_c\bgamma_c}{\gamma_s}\ln(1-\gamma_s)
\right\}
\end{eqnarray}
while the $\beta$ functions of the DOS and $z$ are given by
\bea
\frac{d\nu}{d\ell}&=&g\nu\left\{\frac{1}{2}\ln(1-\gamma_s)+\ln(1-\bgamma_c)-\frac{1}{2}\right\}\\
\frac{dz}{d\ell}&=&-gz\Big(\frac{\gamma_s}{2}+\frac{\bar\gamma_c}{2}+
\frac{1}{2}\Big).
\eea
Solving the equations we find that the RG flow in the ultraviolet limit tend to the line defined by
\be
\gamma_s=\bgamma_c
\ee
at which the interactions become constant under rescaling, namely $d\gamma_s/d\ell=d\bgamma_c/d\ell=0$. 
The final values are generally given by $\gamma_s(\ell\gg 1)\simeq \bgamma_c(\ell\gg 1)
\sim 2\bgamma_c(0)-\gamma_s(0)$. 
In the meantime $g$ goes to strong coupling, if typically $\gamma_s$, $\bgamma_c\ge 0$. 
Notice that in this case, by the transformation $\gamma_s=\bgamma_c\rightarrow -\gamma_t$, one 
obtains the same equations for disordered $d$-wave superconductors with broken time reversal and 
chiral symmetries \cite{npb}. 
We find that, for $\epsilon=0$, there is a critical value for the couplings, $\gamma_s=\bgamma_c\simeq -0.37$, for which
$dg/d\ell=0$, at lowest order in $g$. One, then, should consider second loop corrections. 
However, there are initial values of $\gamma_s$ and $\bgamma_c$, also positive, for which, under RG flow, 
the scattering amplitudes become negative, producing in this way anti-localization effects which suppress 
the resistivity, as clearly 
shown in Fig.~\ref{case4.fig} for a particular choice of initial parameters. 
This means that the interacting system can behave quite differently from the non-interacting one, for which
only a monotonic flow to strong coupling limit is expected.

\subsection{Case 5) class A, with ${\cal S}$ and nonuniform magnetic field (${\cal T}$, SU$(2)$ broken)}

In this case the RG equations are the same as in the last two cases, $9)$ and $10)$ reported below, where 
chiral symmetry ${\cal S}$ is broken. This means that in the presence of broken time reversal symmetry and 
magnetic impurities (or with a nonuniform magnetic field), local random scalar potentials are completely 
irrelevant also in the presence of interactions.    

\subsection{Case 6) class AI, without ${\cal S}$, with ${\cal T}$ and SU$(2)$}
In the absence of chiral symmetry but in the presence of both time reversal symmetry and spin rotational invariance, the $\beta$ functions were derived by Finkel'stein \cite{finkelstein} for long-range (Coulomb) interaction. The full set of equations has been generalized also for short-range interaction \cite{prb} and are the following 
\begin{eqnarray}
\label{(f1)}
&&\hspace{-0.cm}\frac{dg}{d\ell}=-\epsilon g+g^2\left\{5+
\frac{1-\gamma_s}{\gamma_s}\,\ln(1-\gamma_s)
-3\,\frac{1+\gamma_t}{\gamma_t}\,\ln(1+\gamma_t)-\gamma_c
\right\}\\
&&\frac{d\gamma_s}{d\ell}=g\left\{\left(1-\gamma_s\right)\left(\frac{3}{2}\gamma_t+\gamma_c-\frac{1}{2}\right)
+\frac{1}{2}\left(1-\gamma_s\right)^2
\right\}\\
&&\frac{d\gamma_t}{d\ell}=g\left\{\left(1+\gamma_t\right)\left(\frac{1}{2}\gamma_s-\gamma_c-\frac{1}{2}\right)
+\left(1+\gamma_t\right)^2\left(\frac{1}{2}+2\gamma_c\right)\right\}\\
\label{(f4)}
&&\frac{d\gamma_c}{d\ell}=g\left\{\frac{\gamma_s}{2}+\frac{3}{2}\gamma_t
+\gamma_c\left(\frac{\gamma_s}{2}-\frac{3}{2}\gamma_t+\gamma_c+\ln\left[\frac{(1+\gamma_t)^3}{(1-\gamma_s)}\right]\right)\right\}-\left(\gamma_c\right)^2
\end{eqnarray}
For completeness we include also the $\beta$ function for the DOS which is given by
\bea
\label{dosAI}
\frac{d\nu}{d\ell}&=&g\nu\left\{\frac{1}{2}\ln(1-\gamma_s)+ 
\frac{3}{2}\ln(1+\gamma_t)+\gamma_c
\right\}\\
\frac{dz}{d\ell}&=&gz\left(\frac{3\gamma_t}{2}-\frac{\gamma_s}{2}
+\gamma_c\right).
\eea
The long-range interacting version can be obtainded in the limit $\gamma_s=1$ \cite{finkelstein,castellani}. 
\change{Notice that in Eq.~(\ref{(f4)}) and in Eq.~(\ref{dosAI}) the long range limit, $\gamma_s\rightarrow 1$, is divergent. Actually in that limit the DOS has 
double-logarithmic corrections \cite{finkelstein} which can be cured by replacing the running value for the resistance $g$. Similarly, it is argued that such divergence appears in the course of renormalization of $\gamma_c$ (from the term $\langle S_I^{(1)}S_I^{(2)}\rangle$, see Appendix), 
but only at intermediate steps \cite{burmistrov2}.} 
This symmetry class, AI, has been studied \change{at one-loop level to lowest orders in the interaction amplitudes \cite{burmistrov},  
extended to all orders in the interactions $\gamma_s$ and $\gamma_t$ \cite{prb}, in the present form, and completed to all orders also in the interaction $\gamma_c$ \cite{burmistrov2}}, especially 
when the system is close to superconducting instability, finding that 
disorder can promote superconductivity until the system enters the Anderson insulating phase.

\subsection{Case 7) class A, without ${\cal S}$ and ${\cal T}$, with SU$(2)$}
In the absence of both chiral symmetry, as in the case of on-site disorder, and time reversal symmetry, but with preserved spin rotational invariance, the RG equations read
\begin{eqnarray}
\frac{d g}{ d\ell} &=&-\epsilon g+g^2\left\{4-3\frac{1+\gamma_t}
{\gamma_t}\ln(1+\gamma_t)
+\frac{1-\gamma_s}{\gamma_s}\ln(1-\gamma_s)\right\}\\
\frac{d\gamma_t}{d\ell}&=&g\left\{
\left(1+\gamma_t\right)\left(-\frac{1}{2}
+\frac{\gamma_s}{2}
\right)
+\frac{1}{2}  
\left(1+\gamma_t\right)^2
\right\}\\
\frac{d\gamma_s}{d\ell}&=&g\left\{
\left(1-\gamma_s\right)\left(-\frac{1}{2}
+\frac{3}{2}\gamma_t\right)
+\frac{1}{2} 
\left(1-\gamma_s\right)^2
\right\}
\end{eqnarray}
which describe the scaling behavior of the system, together with the $\beta$ function of the DOS and the $z$ parameter
\bea
\frac{d\nu}{d\ell}&=&g\nu\left\{\frac{1}{2}\ln(1-\gamma_s)+
\frac{3}{2}\ln(1+\gamma_t)\right\}\\
\frac{dz}{d\ell}&=&gz\left(\frac{3\gamma_t}{2}-\frac{\gamma_s}{2}\right).
\eea
\begin{figure}[h!]
\includegraphics[width=7cm]{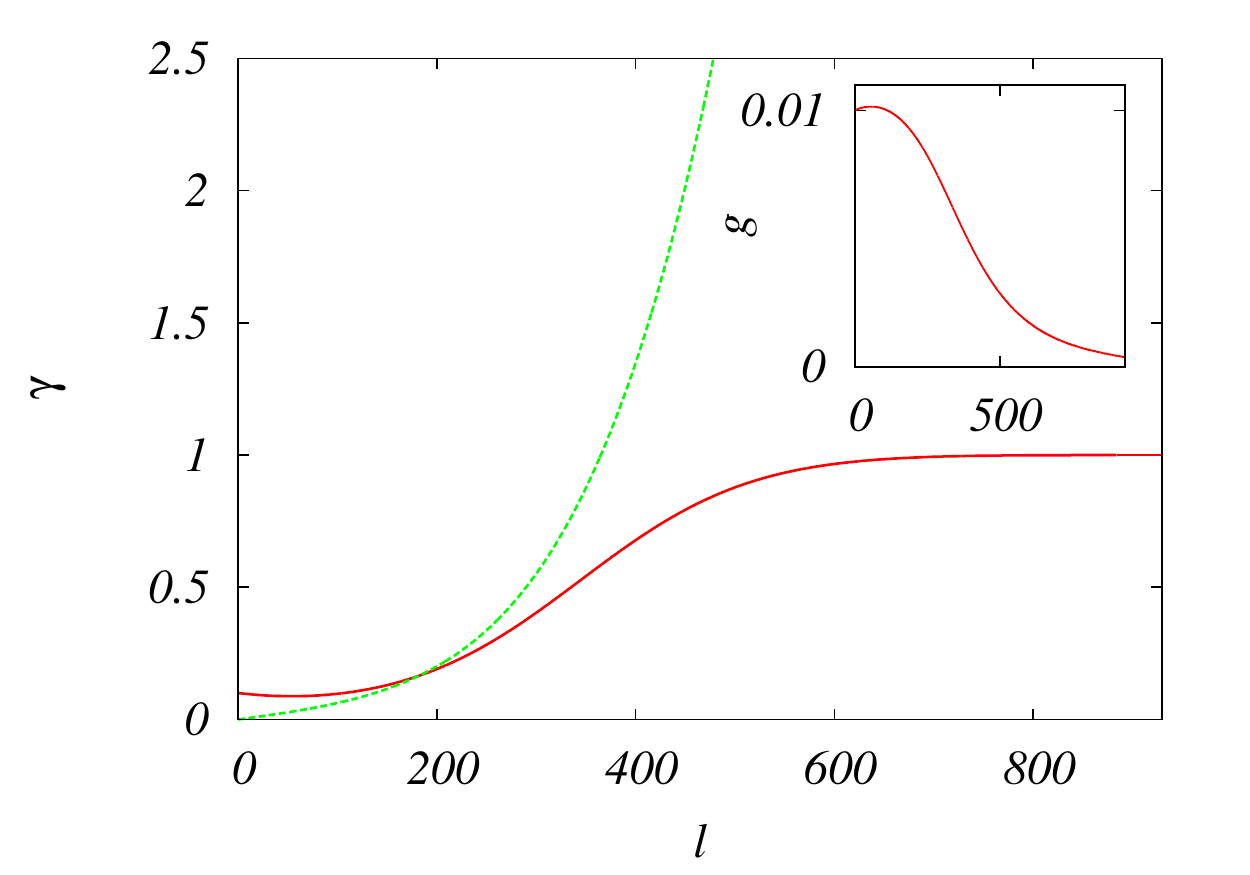}
\caption{Main plot: $\gamma_s(\ell)$ (solid red line) and $\gamma_t(\ell)$ (dashed green line)
for class A, case 7), as functions of $\ell$, the scaling parameter,
with initial values $g(0)=0.01$, $\gamma_{s}(0)=0.1$, $\gamma_{t}(0)=0$.
In the inset: $g$ as a function of $\ell$, for the same initial values.}
\label{case7.fig}
\end{figure}
The symmetries in this case are the standard ones expected for the emergence of
 the IQHE. Let us now analyse the scaling properities far from criticality but in the presence of interactions, described by the equations above.\\
Interestingly, as clearly shown in Fig.~\ref{case7.fig}, under rescaling the system goes towards long-range 
Coulomb interaction which is given by fixing $\gamma_s=1$ \cite{finkelstein,castellani}. 
This value, here, is not fixed from the beginning but is reached under the RG flow. 
At the same time the resistance goes to zero under the effect of an increasing strength of the spin fluctuations. 
\change{It is worth mentioning that in this case the parameter $z$ is known up to second order \cite{burmistrov3}.}

\subsection{Case 8) class AIII, without ${\cal S}$ and ${\cal T}$, 
uniform magnetic field (SU$(2)$ broken)}
If we now break the spin rotational symmetry by a Zeeman term, the equations become
\begin{eqnarray}
\label{gAIIIcase8}
\frac{d g}{ d\ell}
&=&-\epsilon g+
g^2\left\{2+\frac{1-\gamma_s}{\gamma_s}\ln(1-\gamma_s)
-\frac{1+\gamma_t}{\gamma_t}\ln(1+\gamma_t)
\right\}\\
\frac{d\gamma_t}{d\ell}&=&g\left\{\frac{1}{2}
\left(1+\gamma_t\right)\left(1+\gamma_s
\right)-\frac{1}{2}\left(1+\gamma_t\right)^2
\right\}\\
\frac{d\gamma_s}{d\ell}&=&g\left\{\frac{1}{2}
\left(1-\gamma_s\right)\left(\gamma_t
-1\right)
+\frac{1}{2}\left(1-\gamma_s\right)^2
\right\}
\end{eqnarray}
and the $\beta$ functions of the DOS and $z$ are given by
\bea
\frac{d\nu}{d\ell}&=&g\nu\left\{\frac{1}{2}\ln(1-\gamma_s)+\frac{1}{2}\ln(1+\gamma_t)\right\}\\
\frac{dz}{d\ell}&=&gz\left(\frac{\gamma_t}{2}-\frac{\gamma_s}{2}\right).
\eea
\begin{figure}[ht!]
\includegraphics[width=7cm]{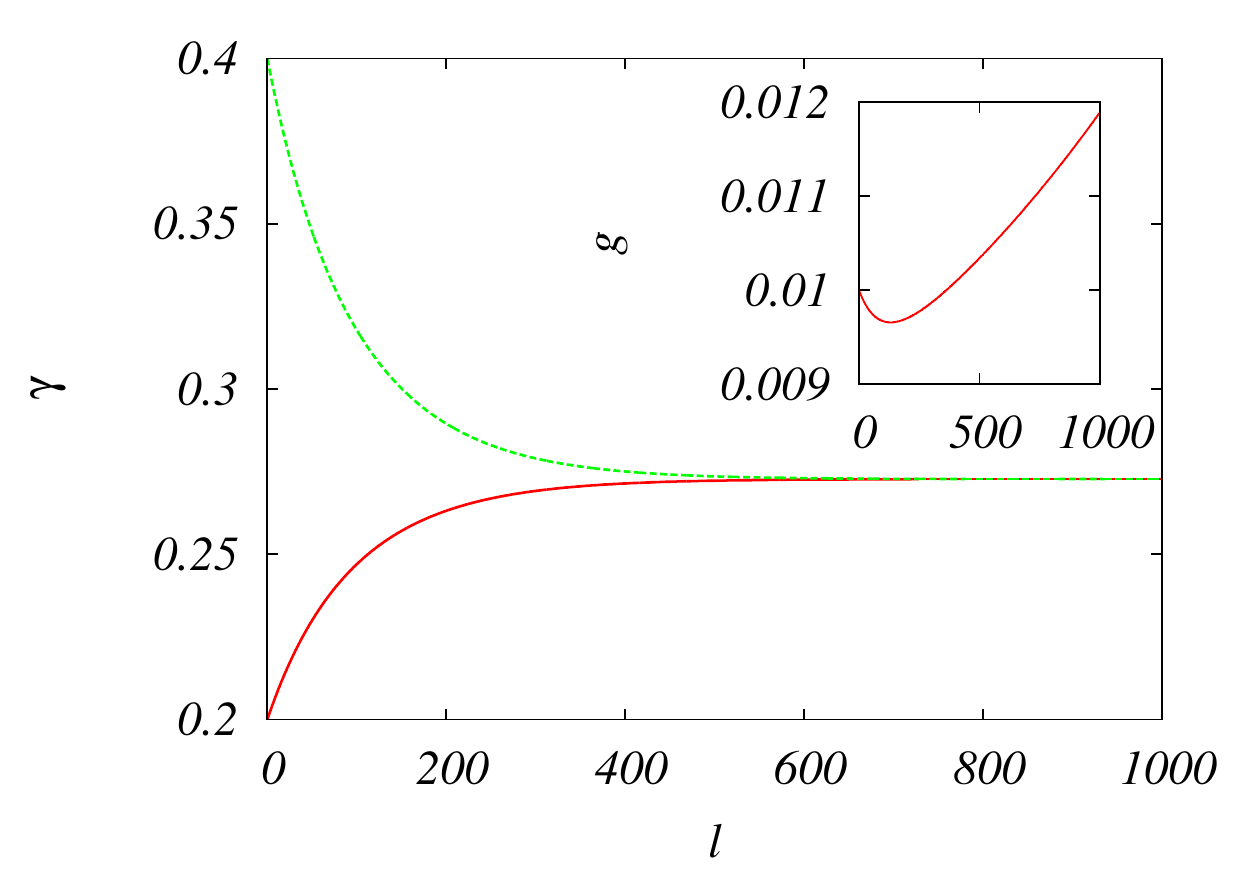}
\caption{Main plot: $\gamma_s(\ell)$ (solid red line) and $\gamma_t(\ell)$ (dashed green line)
for class AIII, case 8), as functions of $\ell$, the scaling parameter,
with initial values $g(0)=0.01$, $\gamma_{s}(0)=0.2$, $\gamma_{t}(0)=0.4$.
In the inset: $g$ as a function of $\ell$, for the same initial values.}
\label{case8.fig}
\end{figure}
In this case, as shown in Fig.~\ref{case8.fig}, under RG, the system tends to evolve towards 
$\gamma_s=\gamma_t$. \\
At that point the interactions do not evolve anymore while the $\beta$ function of $g$, for 
$\gamma_s=\gamma_t>0$, has a positive correction which allows the resistance to flow to strong coupling. 
It is important to remind that either in this case and in case 2) the renormalization of $g$ is 
only due to interactions since in the non-interacting case the resistance is scale invariant, namely 
in the zero replica limit, $dg/d\ell$ has no corrections at any order \cite{gade,wegner}.

\subsection{Cases 9) and 10) class A, without ${\cal S}$,
with or without ${\cal T}$, magnetic impurities (SU$(2)$ broken)}
In the presence of magnetic impurities and with broken chiral sublattice symmetry (as well as in case 5), namely in the presence of chiral symmetry and nonuniform magnetic field) the RG equations are the following
\begin{eqnarray}
\frac{dg}{d\ell}
&=&-\epsilon g+g^2\left\{1
+\frac{1-\gamma_s}{\gamma_s}\ln(1-\gamma_s)
\right\}\\
\frac{d\gamma_s}{d\ell}
&=&
\frac{g}{2}\gamma_s(\gamma_s-1)
\end{eqnarray}
while the scaling behavior of the DOS and $z$ are dictaded by
\bea
\frac{d\nu}{d\ell}&=&\frac{g\nu}{2}\ln(1-\gamma_s)\\
\frac{dz}{d\ell}&=&-gz\frac{\gamma_s}{2}.
\eea
The solution of these equations is a transient since the scattering amplitude $\gamma_s>0$ flows to zero and 
the conductivity changes value only in the meanwhile, reaching a value which, for small interaction, is approximatelly given by
\be
g(\ell\gtrsim g_0^{-1})\simeq g_0\, e^{\gamma_s(0)}\,.
\ee
At the energy scale $\Lambda e^{-1/g_0}$, the resistance $g$ would stop evolving, however the second 
loop correction \cite{hikami,baranov} 
then can contribute driving 
the system to strong coupling, like in the non interacting system.\\ 
It is worthwile stressing that even if the symmetry class is the same, the three cases 3), 7) and 9) have 
different scaling behaviors: case 3) exhibits an antiferromagnetic order, case 7) flows to weak coupling and long-range interacting regime, while in cases 5), 9), 10) the interaction is purly marginal.

\section{Topological terms: strong coupling}

For several classes, the non-linear $\sigma$-model action allows for inclusion 
of topological terms \change{($\theta$ terms,  
which do not appear at any order of the perturbation theory, or Wess-Zumino-Witten (WZW) terms), 
which lead to the emergence of criticality in the strong coupling regime (see for instace \cite{evers})}. 
This occurs, in $d=2$, when the second homotopy group $\pi_2$ of the 
$\sigma$-model manifold ${\cal M}$ ($\pi_d$ is a group of homotopy classes of 
maps of the $d$-sphere $S^d$ into ${\cal M}$) or the third homotopy group 
$\pi_3$, are nontrivial (see Table \ref{Table2}). 
\begin{table}[!h]
\change{
\begin{tabular}{|c||c|c|c|c|c|c|c|c|c|c|}
 \hline
     &\phantom{I}A\phantom{I}&\phantom{.}AI\phantom{I}
& \phantom{.}AII\phantom{.}  &AIII &BDI &\phantom{.}CII\phantom{.}
&\phantom{I}C\phantom{I}   &\phantom{I}CI\phantom{I} &\phantom{I}D\phantom{I}
   &DIII\\\hline\hline
$d=1$&-  &-  &$\Z_2$ &$\Z$  &$\Z$ &$\Z$  &-   &-  &-     &$\Z_2$\\ \hline
$d=2$&$\Z$&-  &$\Z_2$&-     &-   &$\Z_2$ &$\Z$&-  &$\Z$  &-    \\ \hline
$d=3$&-  &-   &-     &$\Z$  &-   &$\Z_2$ &$\Z_2$  &$\Z$&-    &$\Z$\\\hline
\end{tabular}
}
\caption{Homotopy groups of the non-linear $\sigma$-model target spaces (some 
of them listed in Table~\ref{Table}), for all symmetry classes and for 
different dimensions.}
\label{Table2}
\end{table}

In particular, if $\pi_3({\cal M})=\mathbb{Z}$, a WZW term may appear in $\sigma$-models (in $d=2$ for the classes AIII, CI, and DIII), 
\be
S_{WZW}=-i\frac{k}{8}\int_{\Omega} \frac{d R^3}{12\pi}\varepsilon^{\alpha \beta \delta} \Tr\left[\left(Q^\dagger\partial_\alpha Q\right)\left(Q^\dagger\partial_\beta Q\right)\left(Q^\dagger\partial_\delta Q\right)\right]\,,
\ee
where the integration is extended to the third dimension, such that $\Omega$ has as a boundary the real two dimensions. 
This term is the origin of the emergence of criticality at strong coupling limit (large $g$). 
Including this term the topological number {\it k} enters into the renormalization equations only for the disorder parameters $g$ and $\Gamma$, as shown in \cite{xie}, so that, close to the critical point, 
Eqs.~(\ref{gAIII}),~(\ref{GammaAIII}) in case 2), for $\epsilon=0$, should be modified according to 
\bea
\label{gk}
&&\frac{d g_k}{d\ell}=g_k^2\left(1-(kg_k)^2\right)\frac{1}{g^2}\frac{dg}{d\ell}\\
&&\frac{d\Gamma}{d\ell}=g_k\left(1-(kg_k)^2\right) \left(4 +{\Gamma}
\frac{1}{g^2}\frac{dg}{d\ell}\right)
\eea
where $g_k$ is the new running parameter for the resistance in the presence of the WZW term, while $g$ is the resistance without WZW, whose RG flow is governed by Eq.~(\ref{gAIII}). 
Eq.~(\ref{gAIIIcase8}), for case 8), should also be modified as in Eq.~(\ref{gk}). 
There are, instead, five classes for which $\pi_2({\cal M})$ is nontrivial, namely A, C, D, AII, and CII. In the latter cases, for those which has 
$\pi_2({\cal M})=\Z$ (for classes A, C, D), a topological $\theta$-term may be included to the action,
\be
S_{\theta}=-i\theta W[Q]
\ee
where $W[Q]$ is the winding number of the field configuration $Q(R)$. 
One of the most well-known examples is the Pruisken $\sigma$-model for the integer quantum Hall effect 
\cite{pruisken}, which belongs to class A. For the remainig two classes, AII and CII, instead, 
$\pi_2({\cal M})=\Z_2$, for which $\theta$ can only take the values 0 and $\pi$.

As discussed in the previous Section, case 3) belongs to the same symmetry class A of the IQHE in the non-interacting limit. However, we saw that, 
starting from weak coupling, in the presence of interactions, before 
exploring the strong coupling regime, a N\'eel order is formed, \change{contrary to cases 9) and 10), always in the same class of the IQHE, 
where instead interaction is not relevant}. 
The field theory of class C in case 4) is analogous to Pruisken's theory of the IQHE, and is commoly known 
as spin quantum Hall effect (SQHE) \cite{gruzberg,kagalovsky,senthil}, since it occurs in superconductors 
with broken time reversal but 
preserved spin rotational invariance. We showed in this paper that such a symmetry class occurs also in the presence of chiral and time reversal symmetries, and breaking spin rotational invariance. 
This class allows for the presence of a topological $\theta$-term and, consequently, exhibits a critical 
point at $\theta=(2n+1)$ \cite{senthil,read}  
such that the corresponding flow diagram is expected to have qualitatively the same form of the IQHE.
This should be valid also in the interacting case since the interaction in that case is generally a marginal operator. However, for some sets of initial scattering amplitudes, the strong coupling is hardly reached or completely avoided.

\section{Conclusions}
This paper has the aim of being a compendium of $\beta$ functions of several Finkel'stein non-linear $\sigma$-models derived from the simplest choices of interacting random Hamiltonians. 
We consider ten different cases according to the symmetry properties of the original disordered fermionic 
models and derive the corresponding $\sigma$-models. We show that in many cases the presence of the 
interactions modifies qualitatively and quantitatively the scaling behavior of the couplings involved in the field theories, which are the disorder parameters (the resistance $g$, and the parameter $\Gamma$, associated with staggered density fluctuations) and the scattering amplitudes $\gamma$'s. 
For each case the phase diagram can be very rich, specially in the presence of chiral symmetry which allows 
for the emergence of a magnetic order, and, depending on the choice of the strength of the interactions at high energy scale, the system can or cannot flow to localization insulating regime.
In general terms, we show that the Cartan's classification of the symmetry classes is not enough to uniquely determine the scaling properties of the interacting non-linear $\sigma$-models.
\change{
\subsection{Acknowledgments}
I would like to thank I.S. Burmistrov, I.V. Gorny and A.D. Mirlin for useful discussions. 
\section{Appendix}
\noindent
The transverse modes are parametrized by $U$ so that the fluctuating field is (see Eq.~(\ref{UQU}))
\be
\label{UQU_ap}
Q=\tilde{U}^\dagger Q_{sp}U =C U^tC^t Q_{sp}U
\ee
where $C=i\sigma_y \tau_1$ is the charge conjugation, and $U$ fullfils
\be
\label{chargeconjugation}
CU^tC^t=\rho_1 U^\dagger \rho_1=\rho_2 U^\dagger \rho_2
\ee
The unitary transformation $U$ can be written as
\be
U=e^{W/2}=e^{(W_0\rho_0+W_3\rho_3)/2}
\ee
where $W_{0,3}$ are anti-hermitian operators. Imposing the charge conjugacy invariance through Eq.~(\ref{chargeconjugation}), we get 
\bea
\label{cciw0}
CW_0C^t=-W_0=W_0^\dagger,\\
CW_3C^t=W_3=-W_3^\dagger.
\label{cciw3}
\eea
For $\mu\neq 0$, the unitary transformation reduces to $U=e^{W_0\rho_0/2}$, therefore $Q=U^\dagger Q_{sp}U$.\\
Since $Q_{sp}= \lambda$, where $\lambda_{nm}=\textrm{sign}(\omega_n)\delta_{nm}\equiv\lambda_{n}\delta_{nm}$ \cite{finkelstein,gade,npb,foster,efetov,belitz,fabrizio}, the quantum fluctuations are given by
\bea
\label{Ws0}
&&\{W_0,\lambda\}=W_{0, nm}\lambda_m+\lambda_n W_{0, nm}=0,\\
&&\;[W_3,\lambda ]=W_{3, nm}\lambda_m-\lambda_n W_{3, nm}=0,
\label{Ws3}
\eea
so that Eq.~(\ref{UQU_ap}) can be written, with the explicit space dependence, as $Q(R)=Q_{sp}\,e^{W_0(R)\rho_0+W_3(R)\rho_3}$.
If time reversal invariance is broken 
\be
[\tau_3, W_{0}] =[\tau_3, W_{3}] =0.
\ee
If magnetic impurities are present, then 
\be
[\tau_3\vec{\sigma}, W_{0}] =
\{\tau_3\vec{\sigma}, W_{3}\} =0, 
\ee
while
\be
[\tau_3\vec{\sigma}\cdot\vec{b}, W_{0}] =
\{\tau_3\vec{\sigma}\cdot\vec{b}, W_{3}\} =0, 
\ee
in the presence of a constant magnetic field. 
Finally, as already said, 
if chiral symmetry is broken ($\mu\neq0$), we must set $W_3=0$. \\
For each $W_{0,3}$ we can separate the singlet term from the triplet one in the spin space, writing
\be
W_\alpha = W_{\alpha S} + i\vec{\sigma}\cdot\vec{W}_{\alpha T},\label{Wspin}
\ee
where $\alpha=0,3$. In addition we rewrite $W$ in $\tau$-components (Nambu space)
\begin{eqnarray}
\label{W_s}
W_{\alpha S} &=& W_{\alpha S0}\tau_0 + i\sum_{j=1}^3 W_{\alpha Sj}\tau_j,\\ 
\vec{W}_{\alpha T} &=& \vec{W}_{\alpha T0}\tau_0 + i\sum_{j=1}^3 \vec{W}_{\alpha Tj}\tau_j.
\label{W_t}
\end{eqnarray}
The properties of massless modes in the Matsubara frequency space are the following, having imposed the conditions Eqs.~(\ref{cciw0}), (\ref{cciw3}), (\ref{Ws0}), (\ref{Ws3}), 
\be
2\,W_{\alpha \Sigma j,nm}^{ab}=(\pm)_{\Sigma,j}
\big((-1)^\alpha -\lambda_n\lambda_m\big)
W_{\alpha \Sigma j,nm}^{ab*}\,,
\ee
\be
2\,W_{\alpha \Sigma j,nm}^{ab}=[\pm]_{\Sigma,j}
\big((-1)^\alpha -\lambda_n\lambda_m\big)
W_{\alpha \Sigma j,mn}^{ba}\,,
\ee
where 
$\Sigma=S, T$ (singlet and triplet), $j=0,1,2,3$ denotes the (Nambu) 
$\tau$-components, 
$\lambda_n=\textrm{sign}{(\omega_n)}$, where $\omega_n$ are the Matsubara frequencies, $a$ and $b$ are replica indices, 
while, from Eq.~(\ref{Ws0}), $n$ and $m$ are the indeces of odd Matsubara frequencies with opposite sign for $W_0$ and, 
from Eq.~(\ref{Ws3}), with same sign for $W_3$. 
We denote by $(\pm)_{\Sigma,j}$ the signs related to the real or 
the immaginary matrix elements of $W_0$ and $[\pm]$ the signs for the symmetric or antisymmetric matrix elements, listed here
\begin{table}[!h]
\begin{tabular}{|c||c|c|c|c|c|c|c|c|}
 \hline
$\Sigma$,$j$ &$S,0$&$S,1$&$S,2$&$S,3$&$T,0$&$T,1$&$T,2$&$T,3$\\\hline\hline
$(\pm)_{\Sigma,j}$  & $+ $& $-$& $-$& $+$& $+$& $-$& $-$& $+$\\\hline
$[\pm]_{\Sigma,j}$  & $-$& $-$& $-$& $+$& $+$& $+$& $+$& $-$\\\hline
\end{tabular}
\end{table}
\\
The Gaussian propagators, then, read
\bea
\label{gaussprop}
\left\langle W^{ab}_{\alpha {
      \Sigma}\,j,\,nm}(k)W^{cd}_{\alpha {\Sigma
      }\,j,\,rq}(-k)\right\rangle&=&
\big(1-(-1)^\alpha\lambda_n\lambda_m
\big)
\Big[
(\pm)_{\Sigma,j} (-1)^\alpha D^\alpha_{nm}(k)\big(\delta^{ac}_{nr}\delta^{bd}_{mq}\,[\pm]_{\Sigma,j} (-1)^\alpha 
\delta^{ad}_{nq}\delta^{bc}_{mr}
\big)\Big]
\\\nonumber &&\hspace{-0.5cm} 
+2\pi\Pi\,k^2 D^3_{nn}(k) D^3_{rr}(k)
\delta_{\alpha 3}\delta_{{
    \Sigma}S}\delta_{j 0}\delta^{ab}_{nm}\delta^{cd}_{rq}\,, 
\eea
where 
\bea
&&D^0_{nm}(k)=\frac{1}{4\pi\nu}\frac{1}{Dk^2+z|\omega_n-\omega_m|},
\,\,\,\;\,\,\textrm{with}\,\,\,\lambda_n=-\lambda_m,\\ 
&&D^3_{nm}(k)=\frac{1}{4\pi\nu}\frac{1}{Dk^2+z|\omega_n+\omega_m|},\,\,\,\;
\,\,\textrm{with}\,\,\,\lambda_n=\lambda_m
\eea
The factor $z$ is the frequency renormalization and $D = \sigma/(2\nu)$ the diffusion coefficient.
Let us introduce slow and fast modes in the spirit of Wilson Polyakov 
procedure, 
\be
Q=\tilde{U}^{\dagger}_s Q_f U_s=\tilde{U}^{\dagger}_s\tilde{U}^{\dagger}_f Q_{sp} U_f U_s,
\ee
with 
${Q_{sp}}_{n\,m}= \lambda_n \delta_{nm}$. 
$U_s$ contains only slow momentum fluctuations and 
\be
\label{slowU}
{U_s}_{nm}=\delta_{nm}, \,\,\,\, \textrm{if}\,\,\,\,\, \Lambda/s<|\omega_n|<\Lambda \,\,\,\,\,\textrm{or}\,\,\,\,\,\Lambda/s<|\omega_m|<\Lambda,
\ee
where $\Lambda$ acts as an energy cutoff and $s>1$ is the rescaling factor. 
The massless fast modes satisfy by definition
\be
{W_f}_{nm}(k)=0\,\,\,\, \textrm{if}\,\,\,\,\, \{Dk^2,|\omega_n|,|\omega_m|\}<\Lambda/s.
\ee 
Now let us expand the action, Eqs.~(\ref{NLsM}),~(\ref{inter}),
at the Gaussian level in terms of $W_f$, leaving slow $U_s$ unexpanded. 
For simplicity of notation 
we call $W$ the fast modes $W_f$ and denote by $U$ the slow term $U_s$.
In this way, we have 
\bea
&& S_0[Q]\simeq S_0[Q_s]+S_0[W]+S_0^{(1)}[W,U]+S_0^{(2)}[W,U]+S_0^{(z)}[W,U]\\
&& S_I[Q]\simeq S_I[Q_s]
+S_I^{(1)}[W,U]+S_I^{(2)}[W,U]
\eea
such that we can integrate over the fast modes getting an action which depends only on the slow modes
\bea
S[Q_s]&=&S_0[Q_s]+S_I[Q_s]-\ln \int DW e^{-\left(S_0[W]+S_0^{(1)}[W,U]+S_0^{(2)}[W,U]+S_0^{(z)}[W,U]
+S_I^{(1)}[W,U]+S_I^{(2)}[W,U]\right)}\\
&\equiv&S_0[Q_s]+S_I[Q_s]-\ln \big\langle e^{-\left(S_0^{(1)}+S_0^{(2)}+S_0^{(z)}+S_I^{(1)}+S_I^{(2)}\right)}\big\rangle
\eea
where
\bea
&&\hspace{-0.4cm} S_0^{(1)}= \frac{\pi\sigma}{16} \int d\Rv\,
 \Tr\left[ {\bf A}^{ab}_{n_1n_2}\rho_1 \lambda_{n_2}
\Big({\bf A}_{n_3n_4}^{bc} \lambda_{n_4} {W_{n_4n_5}^{cd}W_{n_5n_1}^{da}}
-W_{n_2n_3}^{bc} {\bf A}_{n_4n_5}^{cd} W_{n_5n_1}^{da}\lambda_{n_1}\Big)
\rho_1\right]
\label{F1} \\
&&\hspace{-0.4cm}  S_0^{(2)}=  \frac{\pi\sigma}{8}
\int d\Rv\, \Tr\left(\vec{\nabla}W_{n_1n_2}^{ab}W_{n_2n_3}^{bc} {\bf A}_{n_3n_1}^{ca}\right).
\label{F2}\\
&&\hspace{-0.4cm}  S_0^{(z)}= \frac{\pi z}{8}\int d\Rv\, \Tr\left(\omega_{n_1} \tilde U^{ab\dagger}_{n_1 n_2} \lambda_{n_2}\, W_{n_2n_3}^{bc}W_{n_3n_4}^{cd} U_{n_4 n_1}^{da}\right) \\
&&\hspace{-0.4cm}  S_I^{(1)}= 
\frac{\pi^2\nu^2}{8}\int^{\prime}\sum\Gamma^{\alpha}_{\beta\eta} \,\tr\left(\tilde{U}_{n_{1}m_{1}}^{\dagger de}\lambda_{m_{1}}{W_{m_{1}m_{2}}^{eg}}U_{m_{2}n_{2}}^{gd}\rho_\alpha\tau _{\beta}\sigma_\eta \right)
\tr\left(\tilde{U}_{n_{3}m_{3}}^{\dagger df}\lambda_{m_{3}}{W_{m_{3}m_{4}}^{fh}}U_{m_{4}n_{4}}^{hd}\rho_\alpha\tau _{\beta}\sigma_\eta \right)
\delta_{{n_{1}}\mp {n_{2}}\pm {n_{3}},{n_{4}}} \\
&&\hspace{-0.4cm}  S_I^{(2)}=  \frac{\pi^2\nu^2}{8}\int^{\prime}\sum\Gamma^{\alpha}_{\beta\eta} \,
\tr\left(\tilde{U}_{n_{1}m_{1}}^{\dagger de}\lambda_{m_{1}}
{W_{m_{1}m_{2}}^{eg}}{W_{m_{2}m_{3}}^{gh}}U_{m_{3}n_{2}}^{hd}\rho_\alpha\tau _{\beta}\sigma_\eta\right)\tr\left(Q^{dd}_{n_{3}n_{4}}\rho_\alpha\tau _{\beta}\sigma_\eta \right)\delta_{{n_{1}}\mp {n_{2}}\pm {n_{3}},{n_{4}}},
\eea
where ${\bf A}=\nabla \tilde U\,\tilde U^{\dagger}$ and, if $\beta=0,3$ and $\eta=0$ then $\Gamma^{\alpha}_{\beta\eta}=\Gamma^{\alpha}_s$ (siglet p-h channel), if $\beta=0,3$ and $\eta=1,2,3$ then $\Gamma^{\alpha}_{\beta\eta}=\Gamma^{\alpha}_t$ (triplet p-h channel), if $\beta=1,2$ and $\eta=0$ then $\Gamma^{\alpha}_{\beta\eta}=\Gamma^{\alpha}_c$ (Cooper channel). \\
One can calculate the quantum corrections at one-loop level, getting the same action with renormalized parameters.   
Before starting the renormalization of the parameters involved we can easily take into account the ladder diagrams substituting the bare scattering amplitudes with
\bea
\Gamma_s^0(k,\omega)=\Gamma_s^0\frac{Dk^2+z|\omega|}{Dk^2+(z-2\nu\Gamma_s^0)|\omega|}\\
\Gamma_t^0(k,\omega)=\Gamma_t^0\frac{Dk^2+z|\omega|}{Dk^2+(z+2\nu\Gamma_t^0)
|\omega|}\\
\Gamma_c^3(k,\omega)=\Gamma_c^3\frac{Dk^2+z|\omega|}{Dk^2+(z-2\nu\Gamma_c^3)
|\omega|}
\eea
which are the algebraic ladder resummations. For Cooper amplitude $\Gamma_c^0$, and the staggered amplitudes $\Gamma_s^3$ and $\Gamma_t^3$, we have, already after summing the first two diagrams in the ladder expansion, a logarithmic 
divergent term which does not depend on $g$. 
The one-loop renormalization of the parameters of the NLSM action can be
obtained applying Eq.~(\ref{gaussprop}) and considering the logarithmic corrections which come from different terms: the 
corrections to the conductance $\sigma$ come from 
\be
\label{dsigma}
\big\langle S_0^{(1)}\big\rangle-\frac{1}{2}\big\langle (S_0^{(2)})^2\big
\rangle+\big\langle S_I^{(1)}\big\rangle-
\big\langle (S_0^{(1)}+S_0^{(2)})S_I^{(1)}\big\rangle
+\frac{1}{2}\big\langle (S_0^{(2)})^2 S_I^{(1)}\big\rangle
\ee
where the first two terms are the standard non-interacting contributions. Notice that expanding $\big\langle S_I^{(1)}\big\rangle$ in terms of slow momenta, and neglecting slow frequencies, produces a term $\Tr({\bf AA})$ (see \cite{finkelstein,npb} for further details). This term is obtained also from other expectation values in Eq.~(\ref{dsigma}), together with $\Tr({\bf A}\lambda\rho_1{\bf A}\lambda\rho_1)$, getting a combination of them that reproduces the gradient term of the action, $2\Tr({\bf A}\lambda\rho_1{\bf A}\lambda\rho_1 -{\bf AA})=\Tr(\nabla Q_s\nabla Q_s^\dagger)$.\\
{The corrections to the frequency parameter $z$ is instead the following}
\be
\big\langle S_0^{(z)}\big\rangle-\big\langle S_0^{(z)}S_I^{(1)}\big\rangle+\big\langle S_I^{(1)}\big\rangle
\ee
In this case the expansion in terms of slow frequencies of $\big\langle S_I^{(1)}\big\rangle$ has to be considered in order to get corrections to $z$ \cite{finkelstein,npb}, which 
cancel exactly high order terms coming from $\big\langle S_0^{(z)}S_I^{(1)}\big\rangle$. \\
Finally corrections to the scattering amplitudes $\nu\Gamma^\alpha_{s,t,c}$ come from the following terms 
\be
\big\langle (S_{I}^{(1)}+S_{I}^{(2)})\big\rangle 
-\frac{1}{2}\big\langle (S_{I}^{(1)}+S_{I}^{(2)})^2 \big\rangle 
+ \frac{1}{2}\big\langle (S^{(1)}_{I})^2 S^{(2)}_{I}\big\rangle 
+ \frac{1}{2}\big\langle S^{(1)}_{I} (S^{(2)}_{I})^2 \big\rangle 
- \frac{1}{4}\big\langle (S^{(1)}_{I} S^{(2)}_{I})^2\big\rangle.
\ee
The quantum corrections for $\nu\Gamma^\alpha_{s,t,c}$ combined with those for $z$ give the scaling behavior for the parameters $\gamma_{s,t,c}$ and $\bar\gamma_{s,t,c}$ defined by Eqs.~(\ref{gamma0}) and (\ref{gamma3}). 
As final observation, one can easily check from all the sets of equations 
reported in the main text, that the quantity $(z-2\nu\Gamma^0_s)=z(1-\gamma_s)$, which is related by Ward identities to the compressibility \cite{castellani}, is not renormalized in the absence of sublattice symmetry \cite{finkelstein,castellani} while has a non-trivial scaling behavior in the presence of sublattice symmetry \cite{npb}, namely 
\be
\frac{d}{d\ell}\,z(1-\gamma_s)={\cal C}
\left(1-\gamma_s\right)^2
\ee
with ${\cal C}$ a function of the staggered parameters, and ${\cal C}=0$ for cases 6), 7), 8), 9), 10) (when chiral symmetry ${\cal S}$ is not preserved).
An analogous equation can be written for $(z+2\nu\Gamma_t^0)$, 
related to the spin susceptibility, in the cases where the triplet particle-hole channel is present, 
\be
\frac{d}{d\ell}\,z(1+\gamma_t)={\cal C}_s\left(1+\gamma_t\right)^2
\ee 
where ${\cal C}_s$ is a function of some parameters and is generally different from zero also in the absence of sublattice symmetry, while ${\cal C}_s=0$ when SU$(2)$ symmetry is broken by a uniform magnetic field, cases 3) and 8).
}

\end{document}